\documentstyle[12pt]{article}    

\author{Gary R. Goldstein\thanks{This work is supported in part by funds  
provided by the U.S. Department of Energy (D.O.E.) 
\#DE-FG02-92ER40702.}\\
Department of Physics\\
Tufts University\\
Medford, MA 02155  USA}
\title{Polarization of inclusively produced $\Lambda_c$ in a QCD based
hybrid model}
\date{
29 July 1999}

\begin{document}   

\maketitle

\begin{abstract}
A hybrid model is presented for hadron polarization that is based on
perturbative QCD subprocesses and the recombination of polarized quarks to
form polarized hadrons. The model, originally applied to polarized
$\Lambda$'s  that were inclusively produced by proton beams, is extended
to include pion beams and polarized $\Lambda_c$'s. The resulting
polarizations are calculated as functions of $x_F$ and $p_T$ for high
energies and are found to be in fair agreement with recent experiments.
\end{abstract}

\newpage

\section{Introduction}

\vspace{5mm}

It has long been known that inclusively produced strange hyperons can have
sizeable polarization~\cite{heller1} over a wide range of energies. Since
the initial discovery
of this phenomenon, many theoretical models have
been proposed to explain various aspects of that polarization data, with
varying success~\cite{models,dharma1,degrand,lund}. Because the hyperon
data is in the region of relatively small
transverse momentum ($p_T\sim 1$ GeV/c), soft QCD effects play a major
role in any theoretical explanation. Several years ago W.G.D. Dharmaratna
and G.R. Goldstein developed a hybrid model for hyperon 
polarization in inclusive reactions~\cite{dharma1}. The model involves
hard scattering at the parton level, gluon fusion, to produce a polarized
s-quark which
then undergoes a soft recombination that, in turn, enhances the
polarization of the hyperon. This scheme provided an explanation for the
characteristic kinematic dependences of the polarization in
$p + p \rightarrow \Lambda + X$. The use of perturbative QCD to produce
the initial polarization for strange quarks, with their low current
or constituent quark mass (compared to $\Lambda_{QCD}$) made the
application of perturbation theory marginal. And the magnitude of the
polarization from the gluon fusion subprocesses was quite small.   

More recently some information on polarization of c-quark baryons is
emerging~\cite{accmor,e791}. In this heavy quark realm the perturbative
contribution is more reliable. Given these circumstances, I have modified
the hybrid model to apply to heavy flavor baryons produced inclusively
from either proton or pion beams. The results are encouraging, as the
following will show.

\section{The hybrid model}

\vspace{5mm}

The original hybrid model incorporates the order $\alpha_s^2$ QCD
perturbative calculation of strange quark polarization. The
interference between tree level and one loop diagrams gives rise to
significant polarization~\cite{dharma2}. Of particular
importance for describing existing data on $p+p\rightarrow \Lambda\uparrow 
+ X$ are one loop diagrams leading to strange quark pair production 
initiated by gluon fusion
\begin{equation} g + g \rightarrow  s\uparrow + \bar{s},
\label{qbarq}
\end{equation}
or quark-antiquark annihilation
\begin{equation} q + \bar{q} \rightarrow s\uparrow + \bar{s}.
\label{qqbar}
\end{equation}
The cross sections for polarized s-quarks (polarized normal to the
production plane) must then be convoluted with the relevent structure
functions for the hadronic beam and target. The inclusive cross section
for hadron + hadron $\rightarrow$ s(polarized up or down) + X is obtained 
thereby. For protons on protons gluon fusion is the more significant
subprocess. 

The hadronization process, by which the polarized s-quark recombines with
a (ud) diquark system to form a $\Lambda$, is
crucial for understanding the subsequent hadron polarization. A simple  
prescription is introduced to ``pull'' or accelerate the negatively
polarized, relatively  
slow s-quark along with a fast moving diquark (resulting from a pp
collision) to form the hadron with particular $x_F$ while preserving 
the s-quark's $p_T$ value. Letting
$x_Q$ be the Feynman x for the s-quark, the simple form 
\begin{equation}
x_F = a + bx_Q
\label{xfeqn}
\end{equation}
is used. Naively, if the s-quark has 1/3 of the final hyperon momentum (in
its infinite momentum frame) and the diquark carries 2/3 of that momentum,
then $a=2/3$ and $b=1$. To fit the $pp\rightarrow \Lambda+X$ data (that  
existed in 1990) at one $x_F$ value, the parameters in eqn.~\ref{xfeqn} 
were chosen to be $a=0.86$ and $b=0.70$, not far from the naive
expectation. This recombination
prescription is similar to the classical dynamical mechanism used in the 
``Thomas precession'' model~\cite{degrand},
which posits that the s-quark needs to be accelerated by a confining
potential or via a ``flux tube''~\cite{lund} at an angle 
to its initial momentum in order to join with the diquark to form the
hyperon. The skewed acceleration gives rise to a spin precession for the
s-quark. That non-zero spin component gives the hyperon its polarization,
since the diquark in the $\Lambda$ hyperon is in a spin 0 state, i.e. all 
of the polarization is carried by the s-quark. In the
Dharmaratna and Goldstein hybrid model, however, the s-quark has acquired
negative polarization already from the hard subprocess before it is
accelerated in the hadronic recombination process. That
``seed'' polarization gets
enhanced by a multiplicative factor $A\simeq 2\pi$ that simulates the
Thomas precession. The hybrid model combines hard
perturbative QCD with a simple model for non-perturbative recombination.

The linear form of eqn.~\ref{xfeqn} maps the s-quark Feynman x region
$[-1,(1-a)/b]$ into the $\Lambda$ $x_F$ region $[(a-b),+1]$. The
p+p$\rightarrow$s-quark
differential cross section, $d^2\sigma/dx_Qdp_T$ is mapped
correspondingly into the p+p$\rightarrow \Lambda$ cross section
$d^2\sigma/dx_Fdp_T$. The measured cross sections for the latter are
known to fall with positive $x_F$ and to fall
precipitously with $p_T$, roughly as 
\begin{equation}
(1-x_F)^{\alpha}e^{-\beta p_T^2}
\label{cross}
\end{equation}
overall~\cite{accmor}, where $\alpha$ and $\beta$ are greater than 1.0
(for $\pi+p\rightarrow\Lambda+X$ $\alpha,\beta\approx3.0$). However, the
directly computed lowest order p+p$\rightarrow$s-quark cross section grows
with $x_Q$ in the region (-1,0) and it falls more gradually with $p_T$
than the exponential in eqn.~\ref{cross}. Hence the more complete
recombination scheme would have to temper the $x_F$ dependence and 
narrow the $p_T$ distribution. This will not affect the polarization
calculation, though, since the individual up or down polarized cross
sections will be alterred in the same way. For a more thorough calculation
this should be done, and work is underway on this point. The polarization
results are the focus of this work.

\begin{figure}
\vspace{3in}        
\includegraphics{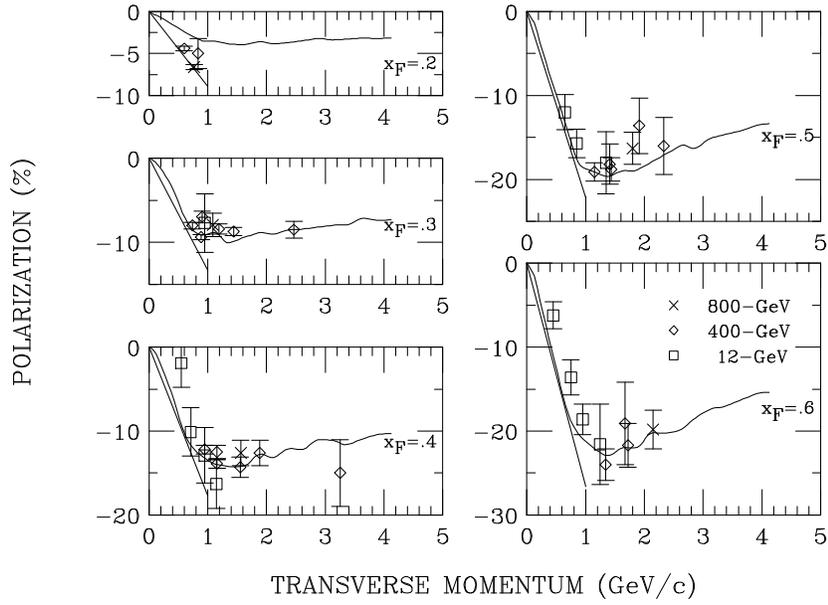}
\vspace{0.5in}
\caption{Hybrid model $\Lambda$ polarization in $p+p\rightarrow \Lambda +
X$ as a
function of $p_T$ for various values of $x_F$. The data at 12
GeV~\cite{heller3}, 400 GeV~\cite{lundberg}, 800 GeV~\cite{ramberg}
are shown. Exclusive data at 27.5 GeV/c~\cite{felix} is approximated by
the straight line from the origin to $p_T\approx 1$ GeV/c.}
\label{fig:gg10} 
\end{figure}

The dominant gluon fusion contribution alone, along with the simple
recombination model, accounted for the contemporaneous polarization
data on $p+p\rightarrow\Lambda+X$~\cite{heller3,lundberg} as well as
subsequent data~\cite{ramberg}. Those data were determined for a range of
energies and $x_F$ and $p_T$ values and the predicted kinematic
dependence was
confirmed, along with the nearly negligible overall energy dependence of
the polarization as fig.~\ref{fig:gg10} shows. Note that the annihilation
subprocess eqn.~\ref{qqbar} has been included, although it makes little
difference for the p+p reaction. The calculated kinematic
dependence is quite unique and was
noticed to be characteristic of hyperon polarization~\cite{heller1}. 

It is noteworthy that extensive data has been collected on $\Lambda$
polarization in several {\it exclusive\/} reactions~\cite{felix}, for
which a simple form, $P = (-0.443\pm 0.037)x_Fp_T$, approximates all the
polarization data at $p_{lab}=27.5$ GeV/c. That form provides lower
bracketting values for the inclusive
polarization, as fig.~\ref{fig:gg10} indicates. In the hybrid model 
all the final states other than the $\Lambda$ arise from the hadronization
of the $\bar{s}$-quark
and the remains of the incoming baryons. Therefore, in the hybrid model it
would be anticipated that as the beam energy increases and/or more final
states are
included in the determination of the $\Lambda$ polarization, more
complicated final states will be accompanied by much lower polarization as
$p_T$ increases beyond 1 GeV/c. 

Since the initial application of the hybrid model, more extensive data on
the polarization of other
inclusively produced hyperons have accumulated~\cite{heller2}. For many of
the reactions some of the
same simple features remain, including the kinematic dependences, while
for others there is a greater complexity. The
ratios of overall polarization for different hyperons are roughly given
by the ratios of their expected SU(6) wavefunctions~\cite{degrand}. The
best measured of these, $\Sigma^+$ polarization~\cite{morelos}, does give
a larger result than -1/3 (the SU(6) factor) times the $\Lambda$
polarization. Comparing the $\Sigma^+$ polarization data of
ref.~\cite{morelos} to the $x_F=0.5$ graph in fig.~\ref{fig:gg10} shows
that the $\Sigma^+$ data is close to -2/3 times the $\Lambda$ data for
all the measured $p_T$ values..
At 800 GeV though, the $\Sigma$ polarization has fallen from
its values at lower energies, suggesting an approach to the lower
value expected from SU(6), but with an energy dependence that is more
pronounced than the $\Lambda$ data.

\begin{figure}
\vspace{3in}        
\includegraphics{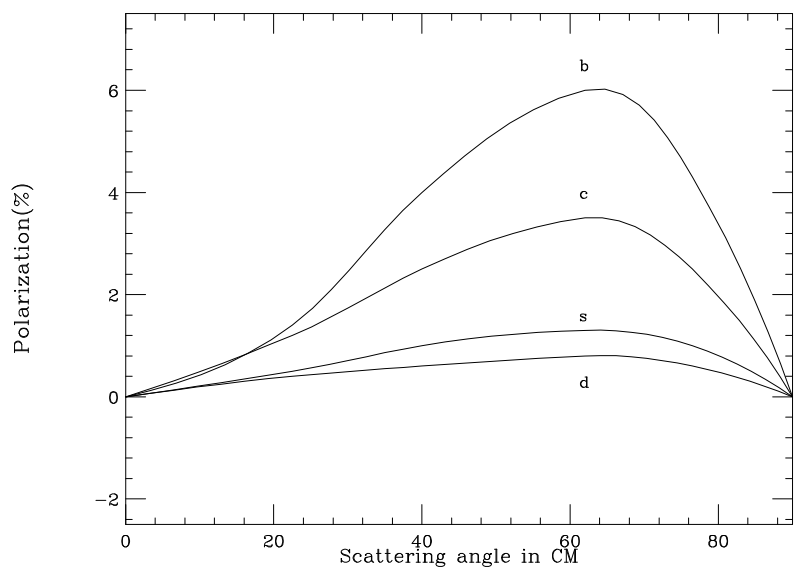}
\vspace{0.5in}
\caption{Polarization for the QCD subprocess of gluon fusion to quark
pairs. The curves are for d, s, c, b quarks.}
\label{fig:gg11}
\end{figure}

In any case, it may be that the PQCD based hybrid
model is best tested in the production of {\it heavy flavor\/} hadrons,
wherein the
heavy quark needs to be produced at large energies compared to the
$\Lambda_{QCD}$ scale. The polarization in the QCD subprocesses was
calculated~\cite{dharma3} to order $\alpha_S^2$ for all  
flavors and it was found that, indeed, the polarization increases
substantially with constituent mass; the peak polarization goes roughly as
the mass, as fig.~\ref{fig:gg11} shows. 

The question then arises, how should the heavy quark recombine with light 
quarks or diquarks to form the heavy flavor hadron? The simplest course to 
take is to adopt the same algebric scheme as for the $\Lambda$ 
hybrid model. The heavy quark again needs to be accelerated by a 
string or flux tube potential and polarization is enhanced via the Thomas 
precession. In heavy quark effective theory, such a polarization
enhancement would arise from the excitation of light degrees of freedom
which contain orbital angular momentum. In any case, the
heavy quark is already polarized when it gets accelerated, so that its
``seed'' polarization, which grows with
mass, is enhanced. This interpretation predicts an increasing polarization 
with increasing quark mass for a given class of final baryons.  

I will now apply the hybrid model to $\Lambda_c$ polarization.
All of the above reasoning was developed with proton-proton scattering in  
mind. It is known that $\pi$ induced inclusive strange $\Lambda$
production also produces significant polarization of $-28\pm10\%$ for
positive $x_F$ and $p_T$ values near
and  above 1~GeV/c~\cite{accmor}. The same experiment found polarized
$\Lambda_c$, with a rough average value over the full range of kinematics
of $\approx -60\%$. Because the beam is a pion, there is a much higher
probability for finding light antiquarks in the beam than for the
proton case. So the quark-antiquark annihilation will be a more important
QCD subprocess for the pion beam than for the proton beam. Both
annihilation and gluon fusion will have to be included. Then in the
recombination the ud-diquark system
that combines with the strange or heavy flavor polarized quark must be
pulled from the target proton or  
the pion sea. In both cases the diquark could have lower average $x_F$
than in the proton beam situation. This might imply a different form for
the Thomas precession enhancement. Furthermore, it should be anticipated
that the mass of the heavy quark will play a role in enhancing the  
polarization further. Various possibilities were considered in determining the
polarization for $\Lambda_c$ via $\pi$ production, where new data
exists~\cite{e791}. Without a thorough exploration of the
recombination parameterization a preliminary
calculation of the pion induced reactions can be made
by adopting the same recombination prescription eqn.~\ref{xfeqn}
used for the $p+p \rightarrow \Lambda + X$.

To summarize, the application of the hybrid model to  estimate
the expected flavor Q baryon $\Lambda_Q$  polarization involves several
steps. First assume $g(x_1)g(x_2)$ and $q(x_1)\bar{q}(x_2) \rightarrow
Q(x_Q,p_T)\bar{Q}$ dominate the
subprocesses that give rise to Q-quark polarization and evaluate the 
polarized cross sections to order $\alpha_s^2$,
$d^2\sigma(\uparrow {\mathrm and} \downarrow)/dx_Qdp_T$. The lengthy
expressions for these subprocess polarizations can be found in  
ref.~\cite{dharma3} (an alternative derivation is given in
ref.~\cite{brandenburg}). Then convolute the polarized subprocess cross
sections with the gluon, quark and antiquark structure functions for
the proton and pion~\cite{duke}, $g^{p,\pi}(x), q^{p,\pi}(x),
\bar{q}^{p,\pi}(x)$, or generically $f_i^{p,\pi}(x)$ leading to 
\begin{equation}
d^2\sigma(\uparrow {\mathrm and} \downarrow)/dx_Qdp_T =
\sum_{i,j}\int_0^1 dx_1\int_0^1 dx_2 f_i^{p,\pi}(x_1)f_j^p(x_2)
d^2\sigma(\uparrow {\mathrm and} \downarrow)/dx_Qdp_T.
\label{dsigma}
\end{equation} 
Next the recombination formula, eqn.~\ref{xfeqn}, is applied to obtain
the corresponding $\Lambda_Q$ polarized cross section at $x_F(=a+bx_Q)$
and $p_T$. The polarization is obtained via
\begin{equation}
P_{\Lambda_Q}(x_F,p_T) =
A\frac{d^2\sigma(\uparrow) - d^2\sigma(\downarrow)}{d^2\sigma(\uparrow) 
+ d^2\sigma(\downarrow)},
\label{polzn}
\end{equation}
in an obvious notation. The parameters a and b  are given the same values
as the original pp scattering case. Regarding A it should be noted,
however, that in the $\Lambda$ data,
for which these parameters were obtained, 20\% to 30\% of the $\Lambda$'s
came from radiative decays of the $\Sigma$. Hence the actual direct
$\Lambda$ polarization should be increased by about 14\%, which increases
the value of A correspondingly to 7.9.
\begin{figure}
\vspace{3in}        
\includegraphics{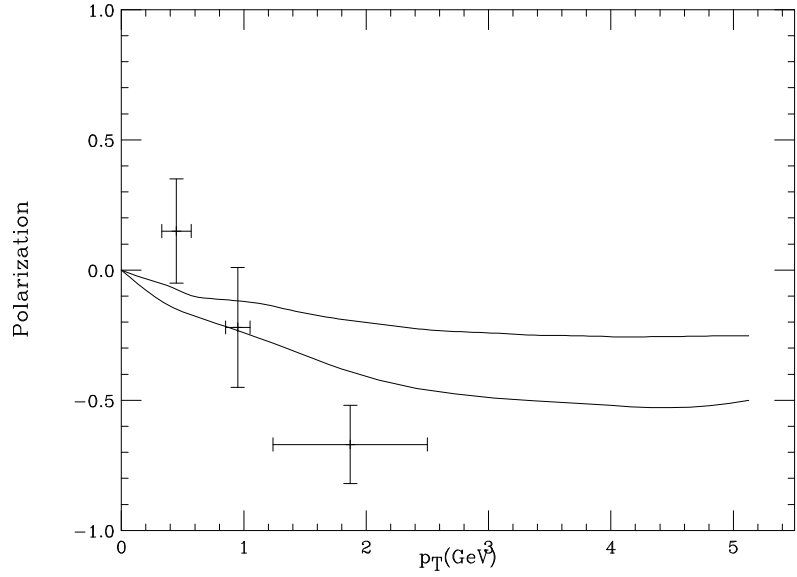}
\vspace{0.5in}
\caption{Estimate of $\Lambda_c$ polarization from $\pi^{-} p\rightarrow
\Lambda_c + X$. The larger polarization includes heavy mass enhancements.
The preliminary data~\cite{e791} is from E791.}
\label{fig:gg12}
\end{figure}

\begin{figure}
\vspace{3in}        
\includegraphics{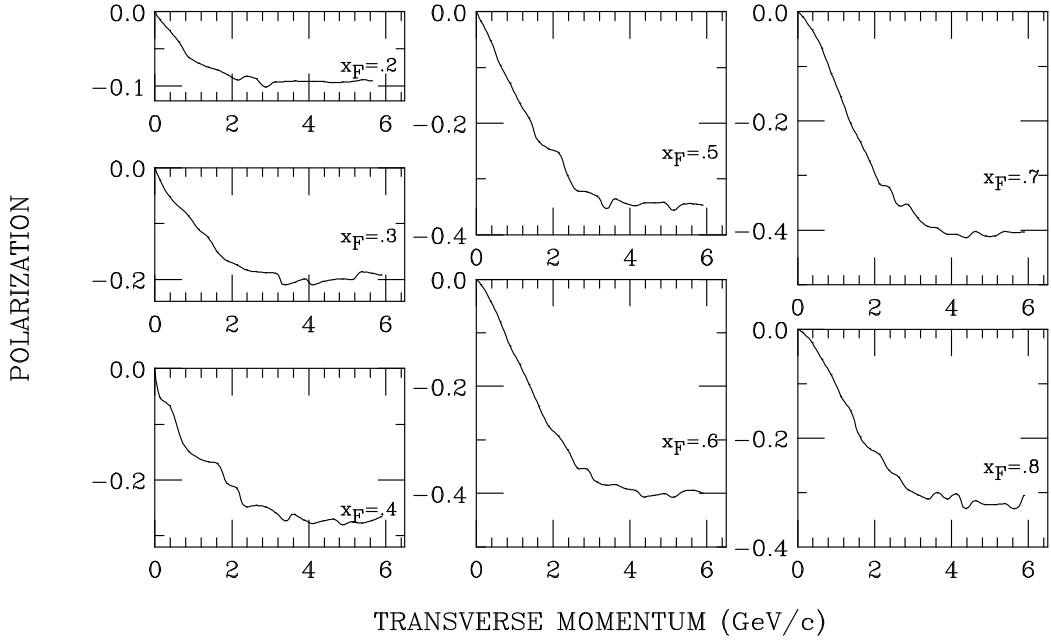}
\vspace{0.5in}
\caption{$\Lambda_c$ polarization in $\pi^- + p\rightarrow
\Lambda_c + X$ as a function of $p_T$ for various values of $x_F$.
Multipying these polarizations by $m(\Lambda_c)/m(\Lambda)$ will
incorporate the hadron mass enhancements as in fig.~\ref{fig:gg12}. }
\label{fig:gg13}
\end{figure}

When this procedure is applied to $\pi^- + p \rightarrow \Lambda_c + X$
the resulting polarization $P(x_F,p_T)$ is obtained. To compare with
preliminary data from Fermilab E791~\cite{e791}, $P$ is integrated over
$x_F$ values from $-0.2$ to $+0.6$~\cite{george}. The resulting $P(p_T)$
is shown as the smaller polarization curve in 
the fig.~\ref{fig:gg12}. To account for the
possible increased enhancement in the Thomas mechanism for the heavier
quark to be accelerated to form the hadron, an additional factor of  
$m_{\Lambda_c}/m_{\Lambda} \simeq 2$ can be included as suggested by
formal studies of the scale dependence on hadronic mass of relevent
correlation functions~\cite{mass}.  With this latter factor,
the larger polarization curve is obtained. This provides a better  
approximation to the new data. It also is consistent with the results for
strange $\Lambda$ production when scaled down appropriately. 
The full dependence on both variables $x_F, p_T$ is shown in
fig.~\ref{fig:gg13}. The polarization corresponds to the smaller
polarization in fig.~\ref{fig:gg12}. It will be of considerable interest  
for the hybrid model to see how well this detailed behavior will be
confirmed when more data are available.

In conclusion, these results are encouraging for the hybrid model.
The Thomas enhanced gluon fusion  
model has been modified to include quark-anti-quark annihilation, which
should be more prominent for heavy baryon polarization in pion induced
reactions, like the above $\pi^- + p \rightarrow \Lambda_c + X$.
Experimental data can be analyzed into $x_F$ as well as $p_T$ bins, so the
predictions from the hybrid model can be checked in detail. The somewhat
{\it ad hoc\/} prescription for the recombination is being studied further
in order to accommodate both the polarization and the cross section
behavior with the kinematic variables. Furthermore,
an investigation of other reactions and observables is underway. 

\section*{Acknowledgments}
This work was supported, in part by a grant from the US Department of
Energy. The author thanks Austin Napier for rekindling a long term
interest in hyperon polarization. He also appreciates correspondence with
members of E791, particularly M.V. Purohit, G.F. Fox and J.A. Appel.

\end{document}